\documentclass[12pt]{article}
\usepackage[height=22cm, width=15.6cm, hmarginratio={1:1}]{geometry}
\title{A Remark on Equivalence between Two Formulas of the Two-point Witten-Kontsevich Correlators}
\author{Jindong Guo}
\date{}
\usepackage{amsthm}
\usepackage{amsmath}
\begin{document}

\maketitle

\begin{abstract}
We prove the equivalence between two explicit expressions for two-point Witten-Kontsevich correlators  obtained by M.~Bertola, B.~Dubrovin, D.~Yang~\cite{BDY} and by P.~Zograf~\cite{Z}.
\end{abstract}

Let $\overline{\mathcal{M}}_{g,n}$ denote the Deligne–Mumford moduli space of stable curves of genus $g$ with $n$ marked points. The genus $g$ intersection numbers of $\psi$-classes   $\langle\tau_{d_1}\tau_{d_2}\dots\tau_{d_n}\rangle_g$ are defined by
	$$\langle\tau_{d_1}\tau_{d_2}\dots\tau_{d_n}\rangle_g=\int_{\overline{\mathcal{M}}_{g,n}}\psi_1^{d_1}\dots\psi_n^{d_n}$$
Here $\psi_i$ denote the first Chern classes of the cotangent line bundles on $\overline{\mathcal{M}}_{g,n}$. 
These numbers, also called the $n$-point Witten-Kontsevich correlators due to their relationship to the KdV hierarchy, vanish unless $d_1+\dots+d_n=3g+n-3$. For $n=2$, M.~Bertola,  B.~Dubrovin and D.~Yang~\cite{BDY} (see page 45 therein)  derives an explicit expression of the two-point Witten-Kontsevich correlators
\begin{equation}\label{formulaBDY}
\langle\tau_{d_1}\tau_{d_2}\rangle_g=\left\{
\begin{array}{lcl}
\frac{\sum\limits_{l=0}^{3g_1+1}{(3g_1+2-l)a_{l-1,3g-l}}}{(6g_1+3)!!(6g_2+3)!!} \quad d_1=3g_1+1, d_2=3g_2+1 \\
\frac{\sum\limits_{l=0}^{3g_1+2}{(3g_1+3-l)a_{l-1,3g-l}}}{(6g_1+5)!!(6g_2+1)!!} \quad d_1=3g_1+2, d_2=3g_2
\end{array}
	\right.
\end{equation}
where $d_1+d_2=3g-1$, and $a_{k_1,k_2}$ is defined by
\begin{equation}\label{coefficientBDY}
	\begin{small}
a_{k_1,k_2}=\left\{
\begin{array}{lcl}
\frac{(6g_1-5)!!(6g_2-5)!!}{2\cdot24^{g_1+g_2-2}\cdot(g_1-1)!(g_2-1)!} &\quad k_1=3g_1-2,k_2=3g_2-2,g_1,g_2\geq1 \\
-\frac{(6g_1-1)!!(6g_2-1)!!}{24^{g_1+g_2}\cdot g_1!g_2!}\frac{6g_2+1}{6g_2-1} &\quad k_1=3g_1,k_2=3g_2-1,g_1,g_2\geq0 \\
-\frac{(6g_1-1)!!(6g_2-1)!!}{24^{g_1+g_2}\cdot g_1!g_2!}\frac{6g_1+1}{6g_1-1} &\quad k_1=3g_1-1,k_2=3g_2,g_1,g_2\geq0\\
0 &\quad otherwise.
\end{array}
\right.
	\end{small}
\end{equation}
Another formula of the two-point Witten-Kontsevich correlators is obtained by P.~Zograf~\cite{Z},  that reads as follows:
\begin{equation}\label{formulaZ}
	\langle\tau_{k}\tau_{3g-1-k}\rangle_g=\frac{(6g-1)!!}{24^gg!(2k+1)!!(6g-1-2k)!!}\biggl(\frac{6g-3}{6g-1}+\sum\limits_{i=1}^{k-1}b_{g,i}\biggr)
\end{equation}
where $b_{g,k}$ is defined by
\begin{equation}\label{coefficientZ}
b_{g,k}=\frac{(6g-3-2k)!!}{(6g-1)!!}\cdot\left\{
\begin{array}{lcl}
\frac{(6j-1)!!(g-1)!}{j!(g-j)!}(g-2j) &\quad k=3j-1 \\	-2\frac{(6j+1)!!(g-1)!}{j!(g-1-j)!} &\quad k=3j \\	2\frac{(6j+3)!!(g-1)!}{j!(g-1-j)!} &\quad k=3j+1
\end{array}
\right.
\end{equation}
\paragraph{}
  Our goal is to prove directly that the two formulas \eqref{formulaBDY} and \eqref{formulaZ} are equivalent. Our proof is based on an identity between the numbers $a_{k_1,k_2}$ and $b_{g,k}$, given by the following lemma.
\newtheorem*{lemma}{Lemma}
\begin{lemma}
The following identity is true:
	\begin{equation}\label{coefficientidentity1}
	\frac{(6g-1)!!}{24^gg!}b_{g,k}=\sum\limits_{l=0}^{k+1}a_{l-1,3g-l}
\end{equation}
\end{lemma}
\begin{proof}
For $k=-1$, the identity is trivial. By direct computation one finds
	 \begin{equation}\label{coefficientidentity2}
	 	\frac{(6g-1)!!}{24^gg!}(b_{g,k}-b_{g,k-1})=a_{k+1,3g-k-2}
	 \end{equation}
 where the computation has three cases according to the value of $k\, ({\rm mod}\,3)$, and by induction the lemma follows.
\end{proof}
Let $k$=$3g_1+1$, $g$=$g_1+g_2+1$ and $k$=$3g_1+2$, $g$=$g_1+g_2+1$, we have
\newtheorem*{theorem}{Theorem}
\begin{theorem}
	The two formulas \eqref{formulaBDY} and \eqref{formulaZ} are equivalent.  
\end{theorem}
\begin{proof}
	The theorem is equal to the following explicit equalities:
\begin{equation}\label{equivalence1}
		\begin{small}
			\begin{aligned}
				\frac{\sum\limits_{l=0}^{3g_1+1}{(3g_1+2-l)a_{l-1,3g-l}}}{(6g_1+3)!!(6g-6g_1-3)!!}&=\frac{(6g-1)!!}{24^gg!(6g_1+3)!!(6g-6g_1-3)!!}\biggl(\frac{6g-3}{6g-1}+\sum\limits_{i=1}^{3g_1}b_{g,i}\biggr)
			\end{aligned}
		\end{small}
	\end{equation}
	\begin{equation}\label{equivalence2}
		\begin{small}
			\begin{aligned}
				\frac{\sum\limits_{l=0}^{3g_1+2}{(3g_1+3-l)a_{l-1,3g-l}}}{(6g_1+5)!!(6g-6g_1-5)!!}&=\frac{(6g-1)!!}{24^gg!(6g_1+5)!!(6g-6g_1-5)!!}\biggl(\frac{6g-3}{6g-1}+\sum\limits_{i=1}^{3g_1+1}b_{g,i}\biggr)
			\end{aligned}
		\end{small}
	\end{equation}
	We begin with proving the first equality \eqref{equivalence1}. By a straightforward calculation, we find that it suffices to prove
\begin{equation}\label{equivalence3}
\sum\limits_{l=0}^{3g_1+1}{(3g_1+2-l)a_{l-1,3g-l}}=\frac{(6g-1)!!}{24^gg!}\sum\limits_{i=-1}^{3g_1}b_{g,i}
\end{equation}
    Observing the following elementary equalities:
$$\sum\limits_{l=0}^{3g_1+1}{(3g_1+2-l)a_{l-1,3g-l}}=\sum\limits_{l=0}^{3g_1+1}\sum\limits_{i=0}^{l}{a_{i-1,3g-i}}=\sum\limits_{l=-1}^{3g_1}\sum\limits_{i=0}^{l+1}{a_{i-1,3g-i}}$$
    and using the Lemma, we find that equality \eqref{equivalence3} is true.
    The proof for the second equality \eqref{equivalence2} is similar.
\end{proof}
In a subsequent publication, we prove in a direct way the equivalence between the $ n $-point functions  considered in this paper \cite{BDY,EO,Zh1,Zh2,Z} and the $ n $-point functions considered by Buryak, Dijkgraaf, Liu--Xu, Okounkov, Zagier (cf.~\cite{B,LX,O}); the explicit relationship between these two types of $ n $-point functions 
can be found in Section 8 of~\cite{BDY}.

\bigskip

\noindent {\bf Acknowledgements.} The author thanks Di~Yang for his advice.

\end{document}